\documentclass[journal=jacsat,manuscript=article]{achemso}

\usepackage[version=3]{mhchem} 
\usepackage{natbib}
\usepackage{graphicx}
\usepackage{amsmath}
\usepackage{multirow,makecell}
\usepackage{bm}
\usepackage{float}
\usepackage{soul}
\usepackage{xcolor}
\usepackage{subcaption}
\usepackage{mciteplus}
\usepackage{siunitx}
\usepackage{adjustbox}
\usepackage{threeparttable}
\usepackage{caption}
\usepackage{indentfirst}
\usepackage{textcomp}

\usepackage{xr}
\usepackage[nameinlink,noabbrev,capitalize]{cleveref}
\SectionNumbersOn

\makeatletter

\newcommand*{\addFileDependency}[1]{
  \typeout{(#1)}
  \@addtofilelist{#1}
  \IfFileExists{#1}{}{\typeout{No file #1.}}
}
\makeatother
\newcommand*{\myexternaldocument}[1]{
    \externaldocument{#1}
    \addFileDependency{#1.tex}
    \addFileDependency{#1.aux}
}
\myexternaldocument{SI}

\author{Xiaoyu Wang}
\affiliation{Department of Chemical and Biomolecular Engineering, University of Notre Dame, Notre Dame, IN 46556, USA}
\author{Yujia Wang}
\affiliation{Department of Chemistry and Biochemsitry, University of Notre Dame, Notre Dame, IN 46556, USA}
\author{Ahmad Moini}
\affiliation{BASF Corporation, Iselin, New Jersey 08830, USA}
\author{Rajamani Gounder}
\affiliation{Charles D. Davidson School of Chemical Engineering, Purdue University, West Lafayette, IN 47907, USA}
\author{Edward J. Maginn}
\affiliation{Department of Chemical and Biomolecular Engineering, University of Notre Dame, Notre Dame, IN 46556, USA}
\author{William F. Schneider}
\affiliation{Department of Chemical and Biomolecular Engineering, University of Notre Dame, Notre Dame, IN 46556, USA}
\alsoaffiliation{Department of Chemistry and Biochemsitry, University of Notre Dame, Notre Dame, IN 46556 USA}
\email{wschneider@nd.edu}

\title
  {Influence of N,N,N-trimethyl-1-adamantyl ammonium (TMAda$^+$) Structure Directing Agent on Al Pair Distributions and Features in Chabazite Zeolite}

\abbreviations{IR,NMR,UV}

\begin{document}

\begin{tocentry}





\includegraphics[width=3.25in]{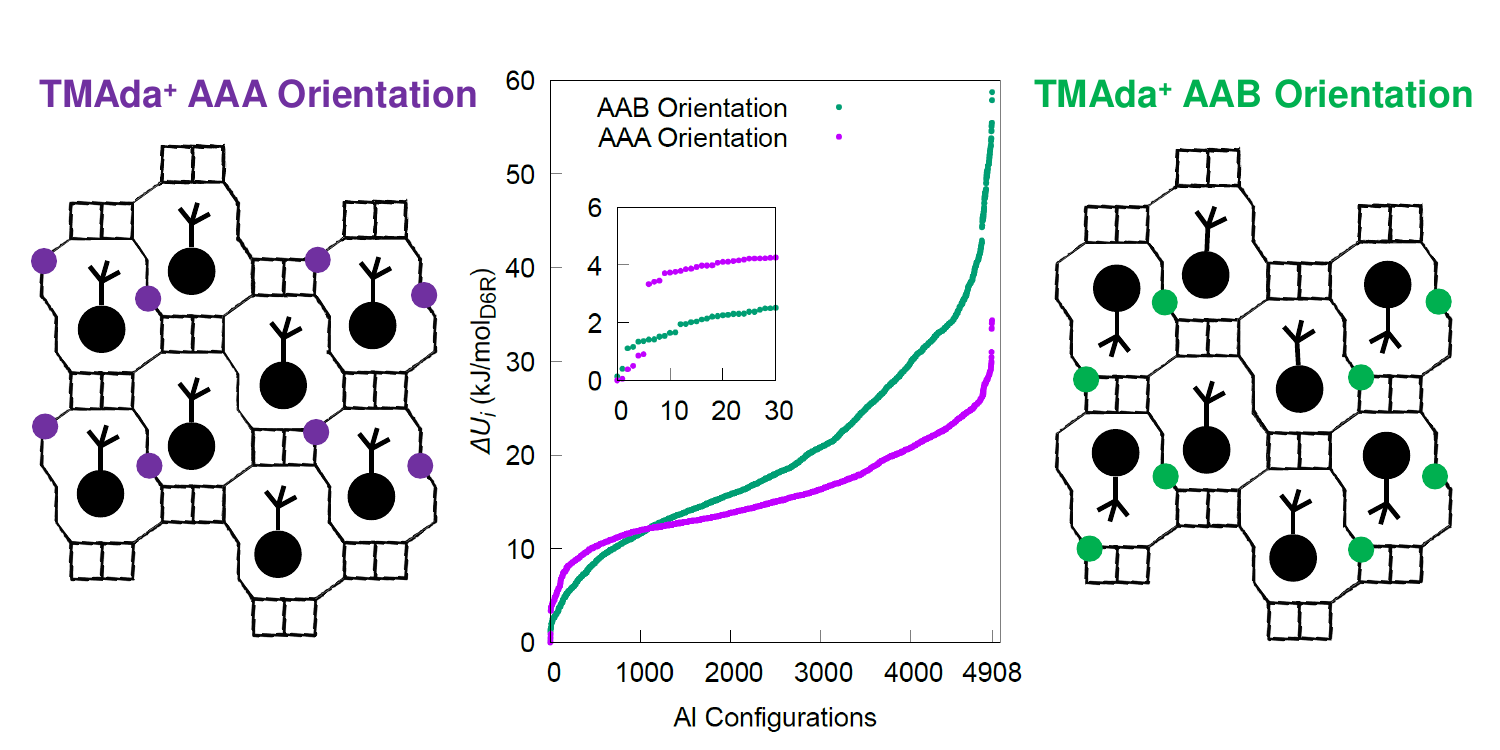}

\end{tocentry}

\begin{abstract} 
While organic structure directing agents (OSDAs) are well known to have a directional influence on the topology of a crystallizing zeolite, the relationship between  OSDA charge and siting of aliovalent ions on a primarily siliceous framework is unclear. Here, we explore the relationship between OSDA orientation, \ce{Al^3+} siting, and lattice energy, taking as a model system CHA zeolite occluded with N,N,N-trimethyl-1-adamantyl ammonium (TMAda$^+$) at an Si/Al ratio of 11/1. We use density functional theory calculations to parametrize a fixed-charge classical model describing van der Waals and electrostatic interactions between framework and OSDA. We enumerate and explore all possible combinations of OSDA orientation and Al location (attending to L{\"o}wenstein's rule) within a 36 T-site supercell. We find that interaction energies vary over 60 kJ/double-six-ring-unit (d6r). Further, analysis of configurations reveals that energies are sensitive to \ce{Al-Al} proximity, such that low energies are associated with \ce{Al^3+} pairs in 8-membered rings and higher energies associated with \ce{Al^3+} pairs in smaller 6- and 4-membered rings. Comparisons with Al siting inferred from CHA zeolite crystallized with TMAda$^+$ suggests that these computed interaction energies are useful reporters of observed Al siting in CHA synthesized with TMAda$^+$.
\end{abstract}


\section{Introduction} \label{introduction}


Zeolites comprise a large class of microporous and crystalline aluminosilicates constructed primarily of silicon-centered and corner-sharing oxygen tetrahedra.\cite{corma2003state}
Specific zeolite topologies are often accessed synthetically through co-crystallization of amorphous phases and gels with organic and/or inorganic structure directing agents (SDAs).\cite{moliner2013towards} 
Organic structure directing agents (OSDAs) are believed to guide zeolites toward particular crystal structures through favorable interactions between the OSDA and the forming framework \cite{gomez2017introduction}.
The computed interaction energy between a preformed framework and an occluded OSDA has been successfully used as a reporter of the potential for an OSDA to crystallize a particular framework. \cite{deem2013computational,schmidt2014synthesis,deem2016computationally} 
This relationship has been exploited to create new zeolites \cite{deem2009computational} and to crystallize zeolites with cages tailored to accommodate the transition state of a target reaction  \cite{boal2016synthesis,gallego2017ab}.

While purely siliceous zeolites are known, the large majority of zeolites contain some amount of aliovalent \ce{Al^3+} substitution onto the \ce{Si^4+} lattice, introducing a net charge onto the zeolite framework. Compensation of that charge by protons generates Br{\o}nsted acid sites useful for various hydrocarbon transformations.\cite{haag1984active}
Further, the relative proximity of those Br{\o}nsted sites within a framework can influence chemical and catalytic properties. \cite{kester2021effects,chen2020tuning,nystrom2018tuning,bhan2008link,roman2011impact}
The \ce{Al^3+} centers can also serve as coordination sites for extra lattice metal ions.\cite{dedecek1994siting,dedecek1995coordination} 
Here too the proximity of centers can have an influence on metal ion speciation, nuclearity, and reactivity.\cite{Walspurger2008109,Giordanino2013,paolucci2016catalysis,li2018first1,li_consequences_2019,devos2019synthesis}

Experimental evidence indicates that synthesis conditions can influence the \ce{Al^3+} siting preferences in zeolites. \cite{knott2018consideration} In zeolites that possess more than one type of symmetry-distinct tetrahedral (T-)site, those conditions can bias \ce{Al^3+} away from or towards particular T-site types, for instance in MFI \cite{yokoi2015control,pashkova2015incorporation,biligetu2017distribution,nimlos2020experimental} and FER \cite{roman-leshkov_impact_2011,pinar_template-controlled_2009}. These effects can be rationalized based on the relative access of charge compensating ions during synthesis to T-sites of distinct environment. On frameworks constructed from a single symmetry-distinct T-site, such as  CHA, these influences are manifested in differences in the proximity of \ce{Al^3+} sites.\cite{di2016controlling,di2017,di2020cooperative,zhang_importance_2020}
CHA is formed from ABC stackings of double-six-ring (d6r) secondary building units and can be crystallized with N,N,N-trimethyl-1-adamantylammonium (\ce{TMAda+}) OSDA \cite{zones1985zeolite}. CHA zeolites crystallized solely with \ce{TMAda+} are observed to exhibit no \ce{Co^2+} uptake capacity and to exchange \ce{Cu^2+} only in its monovalent, \ce{CuOH+} form \cite{di2016controlling},  both indicating that the framework contains no six-membered-rings (6MRs) containing two \ce{Al^3+}. In contrast, synthesis with \ce{Na+} as a secondary, inorganic SDA results in an enrichment in these \ce{Al-Al} pair features in the 6MRs \cite{di2016controlling}. Vibrational spectroscopy provides independent verification of these differences \cite{kester2021effects}.   Synthesis with the larger \ce{K+} cation as the secondary SDA again results in CHA zeolites that lack the 6MR \ce{Al-Al} pair feature \cite{di2020cooperative}.

Charge compensation thus has a determining effect on \ce{Al^3+} siting in CHA. Density functional theory (DFT) calculations show that the proximity-dependent energy of an \ce{Al^3+} pair depends sensitively on the identities of the charge-compensating ions \cite{fletcher2017,li2018first}. Energy is a decreasing function of \ce{Al-Al} separation in the Br{\o}nsted form and exhibits minima at other separations in the presence of mono- (\ce{Na+})\cite{fletcher2017} or divalent (\ce{Cu^2+})\cite{li2018first} cations. The oblong \ce{TMAda+} OSDA is found to occupy the CHA cage in alignment with the long axis; further, the energy of an \ce{Al^3+} charge-compensated by \ce{TMAda+} is a strong function of the separation between \ce{Al^3+} and the charged end of the OSDA, reflecting underlying electrostatic interactions  \cite{li2019influence}. Similar calculations of \ce{TMAda+} and \ce{Na+} or \ce{K+} co-occlusion within the CHA cage are consistent with the former pair resulting in an enrichment in 6MR \ce{Al-Al} pairs while the latter promotes \ce{Al-Al} pairs at greater separation \cite{di2020cooperative}.

\begin{figure}[H]
    \centering
    \includegraphics[scale=0.6]{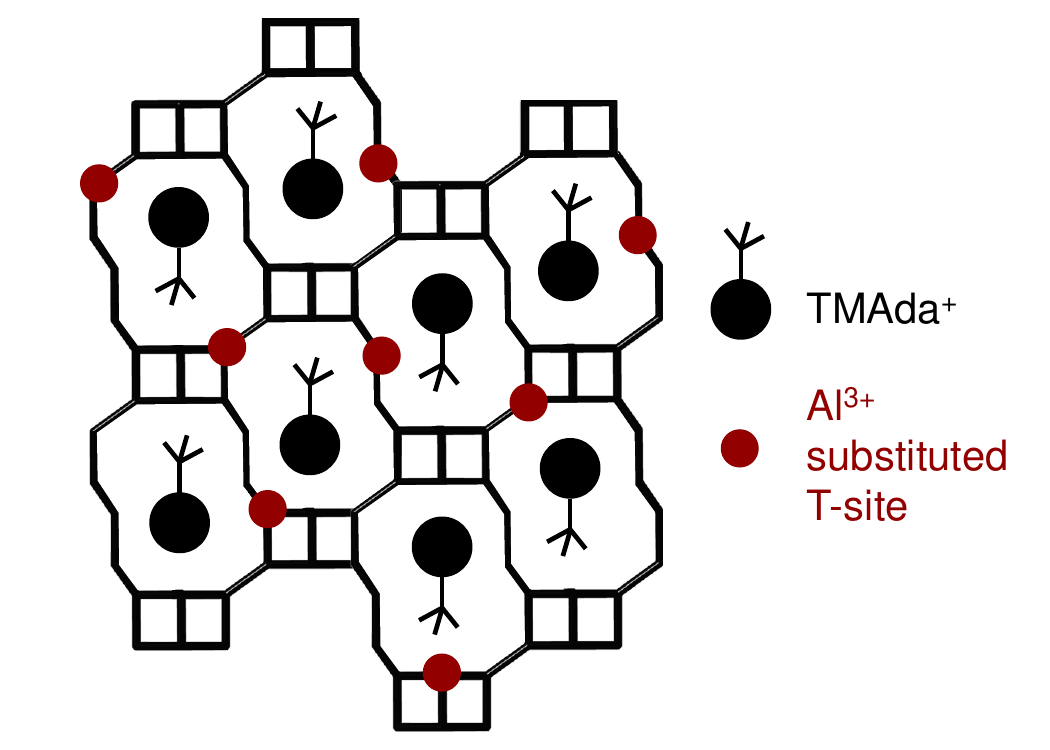}
	\caption{Schematic representation of the CHA/\ce{TMAda+} system. Nodes represent \ce{Si^4+} T-sites, lines bridging oxygen, and red dots \ce{Al^3+} substitutions.Adamantyl body and quaternary nitrogen centers of \ce{TMAda+} represented as black spheres and lines, respectively.}
	\label{system}
\end{figure}

These results highlight the potential to explore the relationship between OSDA and Al siting and proximity more generally. While DFT calculations in principle can provide reliable energy predictions, they are in general too expensive to be used to explore over a wide configuration space of \ce{Al^3+} and OSDA locations. Classical forcefields, however, can be evaluated rapidly and are well suited to capturing the non-bonded and electrostatic interactions most important to the relative energies of \ce{Al^3+} distributions in a field of OSDAs. To explore this approach, we focus here on the CHA/\ce{TMAda+} system.  As shown in the schematic representation in \cref{system}, this system has the advantages of a single, symmetry-distinct T-site and an OSDA that can orient in only one of two equivalent directions within the zeolite cage \cite{li2019influence}. We start with the Dreiding force field successfully applied to neutral analogs of OSDAs \cite{mayo1990dreiding} and augment with charges derived from DFT calculations. We enumerate and explore all possible combinations of OSDA orientation and Al location (attending to L{\"o}wenstein's rule) within a 36 T-site supercell. We find that interaction energies vary over \SI{60}{kJ/d6r}. Further, analysis of configurations reveals that energies are sensitive to \ce{Al-Al} proximity, such that low energies are associated with \ce{Al^3+} pairs in 8-membered rings and higher energies associated with \ce{Al^3+} in smaller 6- and 4-membered rings. Comparisons with Al siting inferred from CHA zeolite crystallized with \ce{TMAda+} suggests that these computed interaction energies are useful reporters of Al siting.

\section{Simulation Details} 

\subsection{DFT and Ab Initio Molecular Dynamics Simulations}
DFT simulations were performed on 36 T-site supercells of varying Al configurations and \ce{TMAda+} orientations using the Vienna Ab initio Simulation Package (VASP), version 5.4.1\cite{VASP}. Lattice constants were obtained from the Database of Zeolite Structures\cite{IZA}. Core-valence interactions were treated using the projector augmented wave (PAW) \cite{PAW,PAW2}, exchange and correlation treated within the Perdew-Burke-Ernzerhof (PBE) generalized gradient approximation (GGA)\cite{PBE}, and the DFT model augmented with the D3 method to describe van der Waals interactions\cite{VDW-D3}.  Plane waves were included to a \SI{400}{eV} cutoff and  the first Brillouin zone  sampled at the Gamma point only. Energies and forces for structures used to parameterize the charge model were converged to \SI{1e-6}{eV} and \SI{0.03}{eV/\angstrom}, respectively (CONTCARs included in the Supporting Information). Single-point calculations were performed on the relaxed structures to generate  AECCAR0, AECCAR2, and CHGCAR files required for performing subsequent atomic population analysis \cite{DDEC-guide}.

Ab initio molecular dynamics (AIMD) simulations were performed at \SI{633}{K} in the canonical (NVT) ensemble, using a Nos\'{e}-Hoover thermostat with Nos\'{e} mass-parameter (SMASS) of $0.01$.  A \SI{1}{fs} time step was used, and hydrogen atoms were replaced by deuterium to accommodate a longer time step. Zeolite framework atoms (Si, Al\textcolor{red}{,} and O) were fixed at positions used in the classical simulations described below during the AIMD simulations to facilitate comparisons with the classical models. At each step, self-consistent-field (SCF) electronic energies were converged to \SI{1e-5}{eV}.  Dynamics simulations were run for \SI{10}{ps} for each configuration. The first \SI{2.5}{ps} of the trajectory was discarded and the remaining \SI{7.5}{ps}  was used to calculate the average potential energy.


\subsection{Classical Force Field Parameterization}
We used the Dreiding force field \cite{mayo1990dreiding}, previously shown to provide good predictions for interactions between OSDAs and siliceous zeolite frameworks \cite{deem2013computational,deem2016computationally}, to describe \ce{TMAda+} and its  van der Waals interactions with  the silica-alumina CHA frameworks considered here.
%
We treated \ce{TMAda+} as flexible and the CHA framework as rigid. Lattice constants and framework atom positions were fixed at those from the Database of Zeolite Structures \cite{IZA}.

\begin{figure} [H]
	\centering
    \includegraphics[scale=0.45]{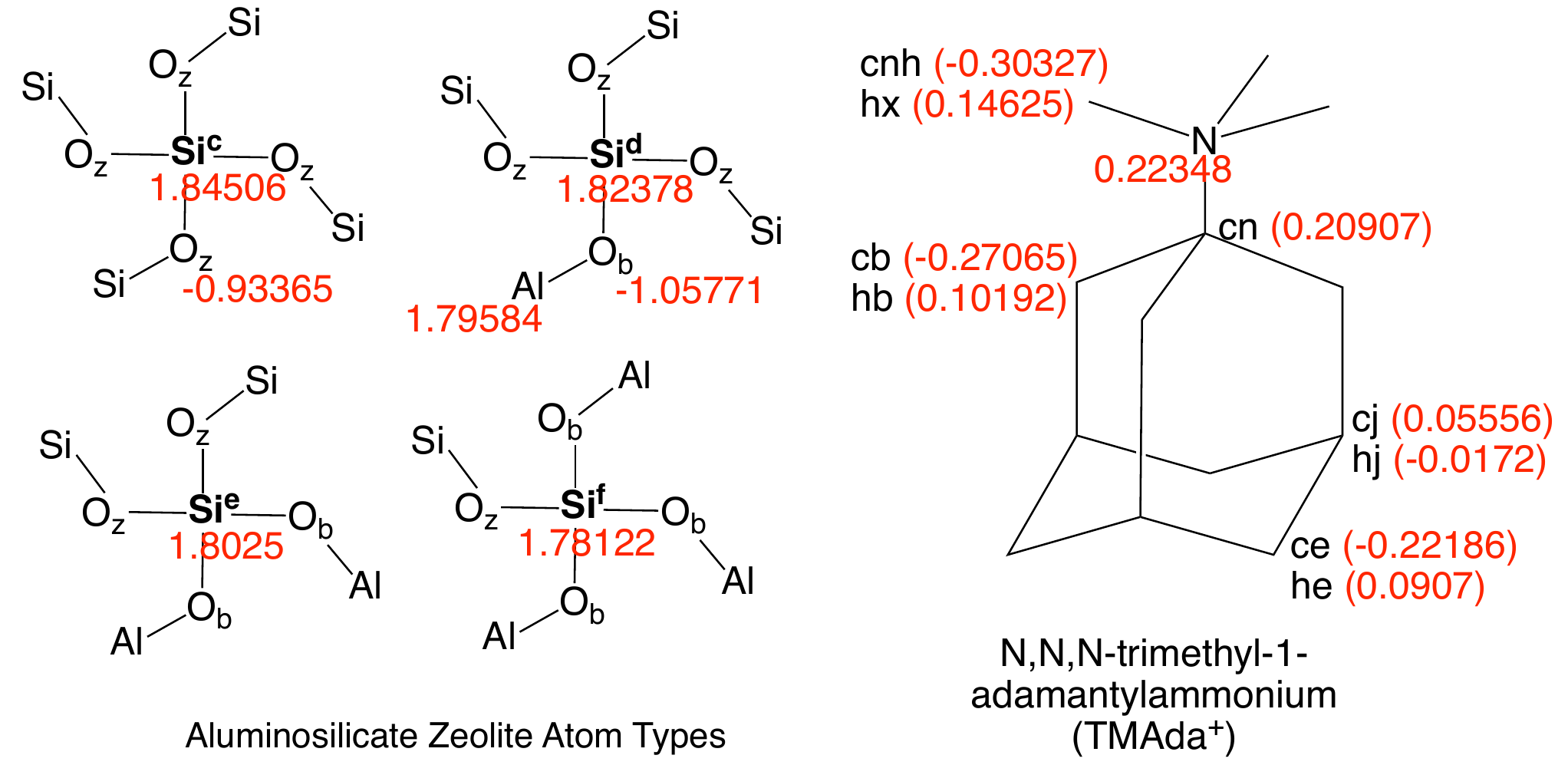}
	\caption{The structure of tetrahedra aluminosilicate unit, TMAda$^+$ and the definition of each atomic type used in \cref{NAC}. Atomic charges are labeled in red beside the atom types.}
	\label{tmada-structure}
\end{figure}

We augmented the Dreiding model with fixed partial charges on \ce{TMAda+} and framework atoms to capture electrostatic interactions.
To derive the partial charges, we choose three arbitrary initial 36 T-site structures and three occluded \ce{TMAda+}, relaxed the structures, and used the Density Derived Electrostatic and Chemical (DDEC) approach \cite{manz2010chemically,manz2012improved,manz2016introducing} to extract atomic net charges. XYZ files containing raw partial charges obtained from the different minimizations are provided as a zipped file in Supporting Information. To reduce the number of distinct atom types, and based on analysis of the DDEC-derived charges, we characterized atoms based on distinct chemical environments (\cref{NAC,tmada-structure}). We reserved one atom type for Al and two atom types for O, including \ce{O_{b}} ions that connect Al and Si and \ce{O_{z}} ions that connect two Si. The Si charge is sensitive to the number of neighboring \ce{AlO4-} tetrahedra, leading to four distinct Si atom types. \ce{TMAda+} C and H atoms are categorized based on their positions relative to the quaternary ammonium group.
Atomic charges were derived by averaging over raw charges from the three configurations and imposing overall electroneutrality.  The procedure was repeated on ten additional relaxed DFT structures and charges found to vary by less than 5\%. The relative energies are also insensitive to the partial charge variations, as two different partial charge sets only lead to ~\SI{1.0}{kJ/mol_{d6r}} energy difference across tested configurations.

\begin{table} [H]
\setlength{\tabcolsep}{10pt} 
\renewcommand{\arraystretch}{1.2} 
\begin{threeparttable}
\begin{tabular}{ccc|ccc}
\hline\hline
Molecule &  Atom type  &  $q$($e$) & Molecule &  Atom type  &  $q$($e$)\\
\hline
CHA zeolite  & Al & 1.79584 & TMAda$^+$  & n & 0.22348 \\
             & O\ce{_b}\tnote{a} & -1.05771 & & cnh & -0.30327\\
             & O\ce{_z}\tnote{b} & -0.93365 & & hx & 0.14625\\
             & Si\tnote{c} & 1.84506 & & cn & 0.20907\\
             & Si\tnote{d} & 1.82378 & & cb & -0.27065\\
             & Si\tnote{e} & 1.8025 & & hb & 0.10192\\
             & Si\tnote{f} & 1.78122 & & cj & 0.05556\\
             & & & & hj & 0.0712 \\
             & & & & ce & -0.22186 \\
             & & & & he & 0.0907 \\
\hline
 \end{tabular}
 \begin{tablenotes}
 \item[a] Oxygen bridging Al and Si.
 \item[b] Oxygen bridging two Si.
 \item[c] Si without first-neighbor Al.
 \item[d] Si with one first-neighbor Al.
 \item[e] Si with two first-neighbor Al.
 \item[f] Si with three first-neighbor Al.
 \end{tablenotes}
 \end{threeparttable}
 \caption{Net atomic charges on zeolite and TMAda$^+$ atoms}
 \label{NAC}
\end{table}

\subsection{Classical Molecular Dynamics Simulations}
The LAMMPS \cite{plimpton1995fast} package was used to carry out all classical molecular dynamics (CMD) simulations.  The cell parameters and locations of atoms, which were obtained from the Database of Zeolite Structures\cite{IZA}, are the same as used in AIMD simulations. 
Each simulation was equilibrated for \SI{500}{ps} followed by a production run of \SI{1500}{ps}, all using a time step of \SI{0.2}{fs}.
The production runs were divided into three sections, from which three block-averaged potential energies and their standard deviations were calculated to reflect the fluctuation of the energies.
The NVT ensemble with the Nos{\'e}-Hoover \cite{nose1984unified,hoover1985canonical} thermostat at \SI{433}{K} was applied.
A cutoff of \SI{10}{\angstrom} was used for non-bonded and electrostatic interactions.
A standard long-range van der Waals tail correction was added to the energy and pressure, while a particle-particle particle-mesh solver \cite{hockney2021computer} was used to describe the long-range electrostatics.

\section{Results and Discussion} 

\subsection{Al Configurations and OSDA Orientations}

\begin{figure}[t]
\centering
\includegraphics[width=0.95\columnwidth]{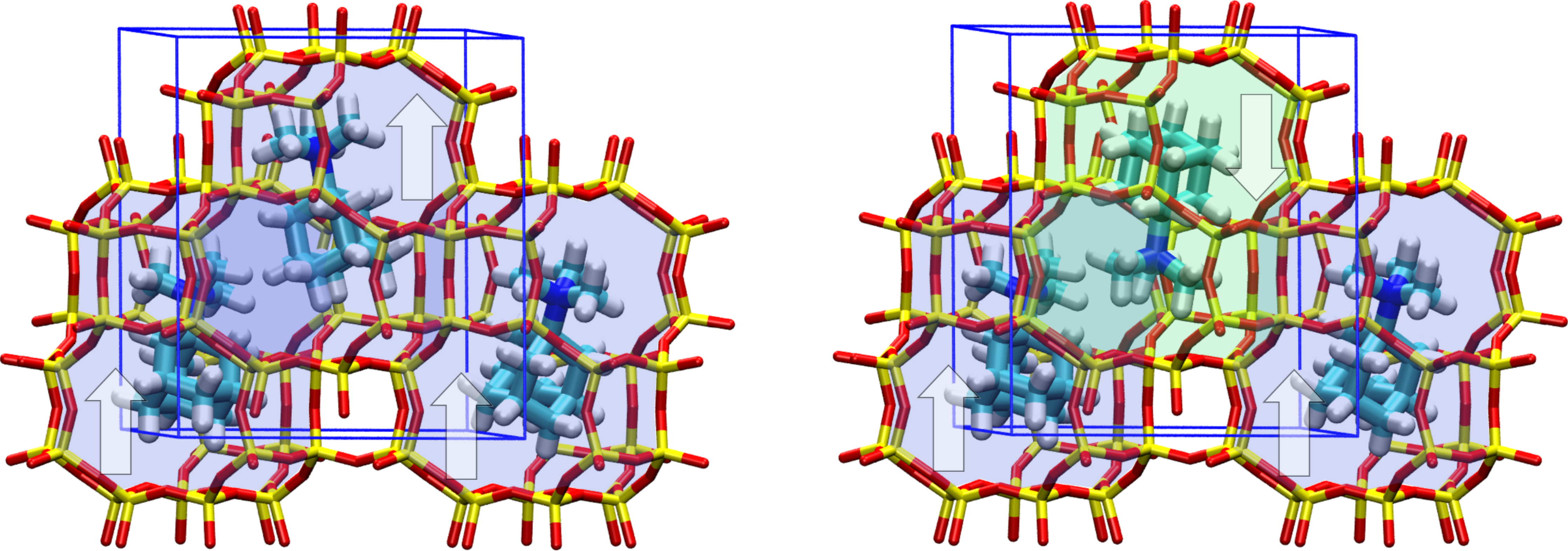}
\caption{36 T-site CHA supercell. Periodic cell boundaries shown in blue, T-sites in yellow, oxygen in red. Left and right images illustrate two unique (``AAA'' and ``AAB'') occlusions of \ce{TMAda+}, orientations highlighted with blue or green shading of cha cages and with arrows.}
\label{aaa-aab}
\end{figure}

We seek to explore the relationship between Al siting, \ce{TMAda+} orientation, and system energy. To that end, we systematically enumerated all combinations of Al location and \ce{TMAda+} orientations possible within a 36 T-site supercell containing three unique cages, shown in \cref{aaa-aab}.  Three Al were distributed over all possible T-sites, excluding those that contain Al-O-Al linkages that violate L{\"o}wenstein's rule \cite{loewenstein1954distribution} and are not properly described by the classical force field. \textcolor{red}{By enumerating all possibilities of Al residing on different T-site locations, we make sure that all possible Al configurations are adequately sampled, which brings us 4908 different Al combinations.} Three \ce{TMAda+} were then introduced to these 4908 configurations  either all in the same orientation (``AAA'') or with two \ce{TMAda+} cations oriented in one direction and the third pointing in the opposite one (``AAB''), as illustrated in \cref{aaa-aab}, resulting in 9816 separate initial configurations that span all possibilities within the 36 T-site supercell. \textcolor{red}{Since the CHA zeolite is a high symmetry framework and the 36 T-site framework is symmetric from three directions, by only considering ``AAA'' and ``AAB'' we can still ensure that all different Al with \ce{TMAda+} configurations have been investigated. All ``AAA'' and ``AAB'' structures have been provided in SI in the ``xyz'' format.}

\begin{figure}[H]
    \centering
    \includegraphics[scale=0.8]{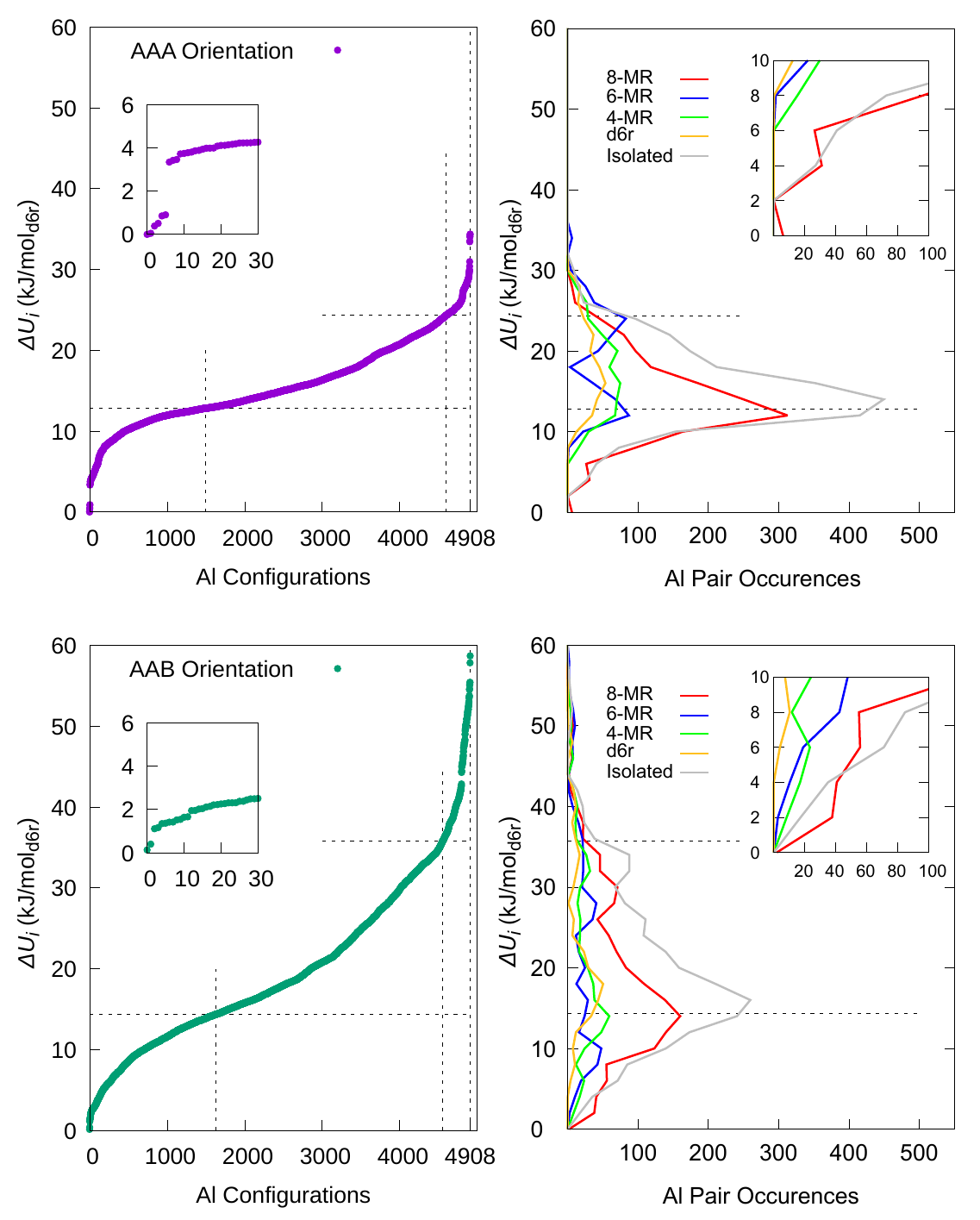}
	\caption{Mean potential energies ($\Delta U_i$) of AAA configurations, sorted in ascending order (top left and inset); each dot represents the energy for an individual configuration. Histograms of \ce{Al-Al} proximity features vs potential energy (top right and inset). Histogram bin size is \SI{2.0}{kJ/mol_{d6r}}; consecutive bins are connected by straight lines. Corresponding results for AAB configurations in bottom left and right. }
	\label{pe_aaa}
\end{figure}

Molecular dynamics simulations were performed at \SI{433}{K} on each of these 9816 configurations. Consistent with prior work \cite{li2019influence}, \ce{TMAda+} maintain their orientation throughout the simulation. Potential energy fluctuations, which were calculated from three block averages, during the MD simulations are less than \SI{5}{kJ/mol_{d6r}} after initial equilibrations of \SI{500}{ps}. The average potential energy $\langle U \rangle_i$ of each configuration $i$ was computed every \SI{1000}{timesteps} and the relative energy of each configuration $\Delta U_i$ was computed as
\begin{equation} \label{delta_e}
\Delta U_i = ( \langle U \rangle_i - U_\text{ref} ) / N_{d6r}
\end{equation}
where the reference potential energy $U_\text{ref}$ is taken as the lowest energy configuration
\begin{equation} \label{e_ref}
U_\text{ref} = \min\limits_{i} \{ \langle U \rangle_i \}
\end{equation}
and the energy is normalized by the number of d6r units. The left panels of \cref{pe_aaa} report \(\Delta U_i\), sorted from lowest to highest energy, for the Al configurations in a field of AAA- and AAB-oriented TMAda$^+$, respectively. Several observations are immediately evident. First, energies span nearly $40$ and \SI{60}{kJ/mol_{d6r}} in the AAA and AAB data sets, respectively, reflecting a substantial sensitivity to Al configuration within a given field of \ce{TMAda+} and different sensitivities to different fields. Second, the lowest energy Al configurations in the AAA and AAB sets are of similar energy. And third, a small handful of configurations dominate the low energy (and high energy) regimes.

\begin{figure}[tb]
\centering
\includegraphics[scale=0.45]{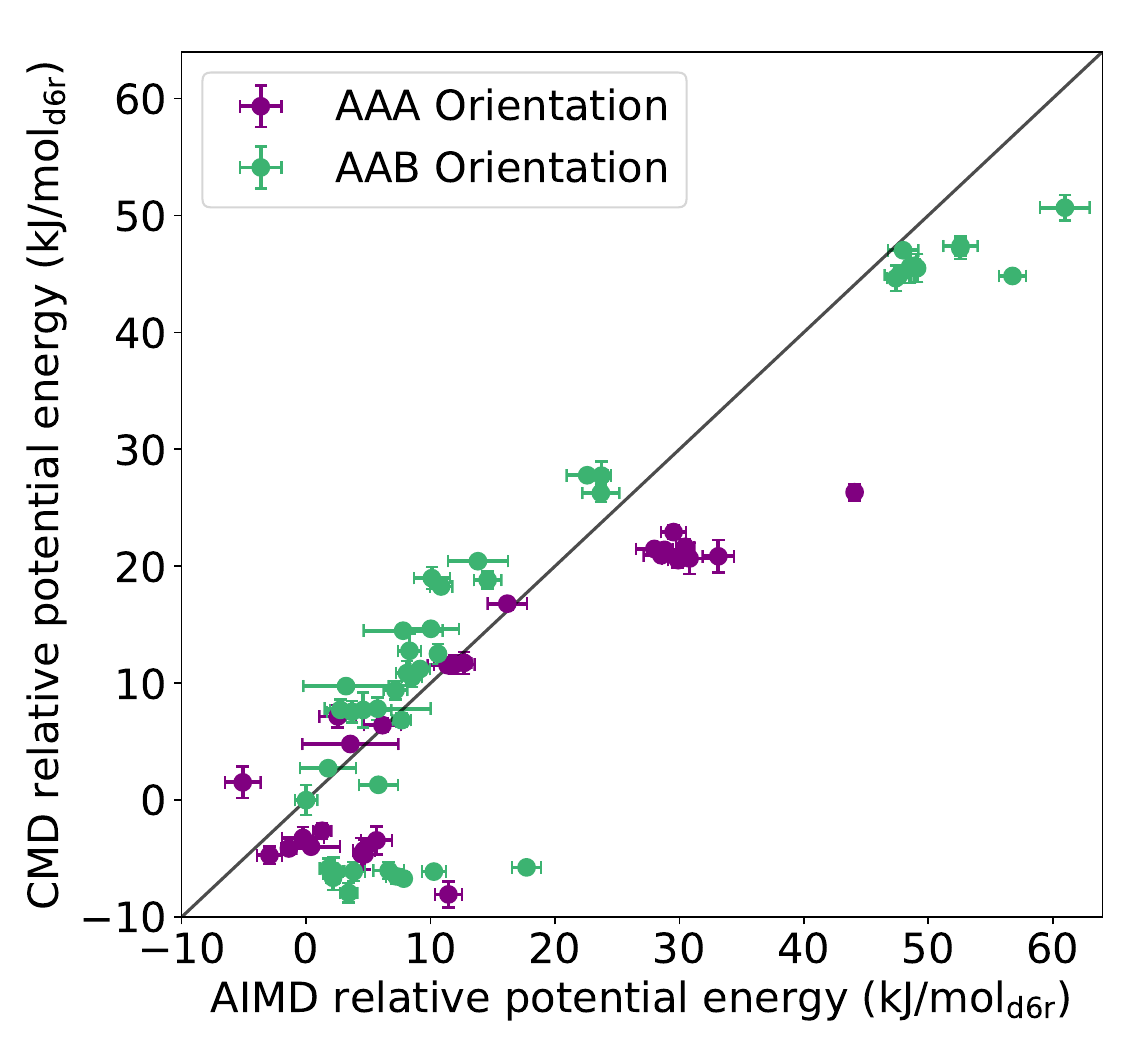}
\caption{Parity plot of CMD potential energies against AIMD potential energies (per d6r unit). Purple and green dots correspond to AAA and AAB arrangements, respectively, and whiskers correspond to error estimates as described in text. }
\label{AIMD_CMD}
\end{figure}

Before analyzing the relationship between configurations and energy, we tested classical model predictions against AIMD results.  We chose the ten lowest- and ten highest-energy configurations from the AAA and AAB sets and augmented with a number of intermediate energy configurations to create a basket of 72 configurations.  Initial structures were extracted from the last frame of the \SI{2}{ns} CMD simulations (POSCARs available in the Supporting Information).  AIMD simulations were run at \SI{633}{K} for \SI{10}{ps} and the last \SI{7.5}{ps} of  trajectory  used to calculate the average potential energy. Uncertainties were taken as the standard deviation obtained from three equal length blocks from the trajectory.  \cref{AIMD_CMD} plots the mean CMD energies against the AIMD energies, color coding according to \ce{TMAda+} orientation, and choosing the lowest energy AIMD AAB configuration as reference. The uncertainties in the AIMD and CMD energies span comparable ranges. Mean energies and uncertainties are listed in Table S1 in Supporting Information.  

The energy range spanned by the AIMD results are consistent with the CMD-predicted spans and differences between the AAA and AAB subsets.  Further, the best fit line through the data has a correlation coefficient of $0.91$, consistent with a robust correlation between the two models. Nonetheless, some substantial differences are evident. Within the envelope of low energy structures, the AIMD energy variations are a factor of four greater than the CMD. Similar but smaller variations are evident in the higher energy envelope. Figure S1 of the Supplementary Information shows the correlation between CMD and AIMD energy differences for 20 structures with lowest CMD energies and the average of the reciprocals of \ce{Al-Al} distances in those structures. The CMD and AIMD energy differences are obtained by $E_{\mathrm{CMD}}$-$E_{\mathrm{AIMD}}$ using data plotted in \cref{AIMD_CMD}. \ce{Al-Al} distances are obtained by calculating the distances among three Al in the supercells considering the minimum-image convention. \textcolor{red}{Since a pair of \ce{Al-Al} can span across periodic images, we used the minimum-image convention to ensure that only the shortest \ce{Al-Al} distance is considered. By applying such a method, we can generate three shortest distances of the three pairs of \ce{Al-Al} in a given CHA framework.} The CMD-AIMD errors correlate with the \ce{Al-Al} separation. The correlation shows CMD overpredicts the stability of the structures that contain short \ce{Al-Al} separations. Structures that contain most 2NN \ce{Al-Al} pairs have the largest errors. These discrepancies may represent limitations of the number of distinct atom types in the classical model. We conclude that the CMD model properly captures the larger energy variations in the system.

\begin{figure}[tbh]
\centering
\includegraphics[scale=0.43]{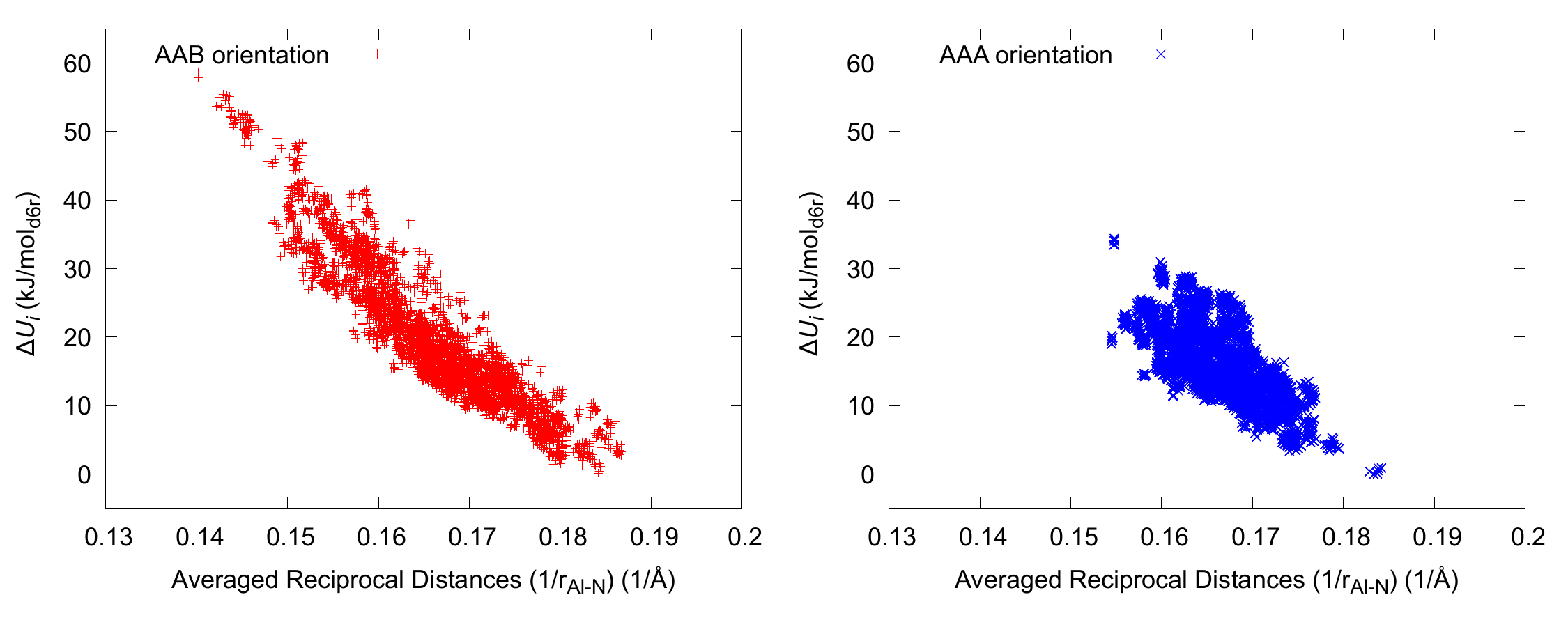}
\caption{Relative potential energies ($\Delta U_i$) vs reciprocal Al to quaternary ammonium N distances ($1/r_{\ce{Al-N}}$) across Al configurations in a field of AAB (left) and AAA (right)  TMAda$^+$.}
\label{pe_vs_dist}
\end{figure}

To explore the relationship between gross structure and energy, and in particular to look for the signatures of an electrostatic contribution to the large energy spans seen in \cref{pe_aaa}, we created parity plots of relative potential energies against ensemble-averaged $1/r_{\ce{N-N}}$ (Supplementary Figure S2), $1/r_{\ce{Al-Al}}$ (Supplementary Figure S3), and $1/r_{\ce{Al-N}}$ (\cref{pe_vs_dist}) across all AAA and AAB configurations. Averages here are over the three shortest minimum image distances, which would be expected to capture the leading electrostatic contributions of each pair-wise interaction.  $\Delta U_i$ is essentially uncorrelated with $1/r_{\ce{N-N}}$ and $1/r_{\ce{Al-Al}}$. In contrast, and as seen in \cref{pe_vs_dist}, energy and $1/r_{\ce{Al-N}}$ are anti-correlated, so that configurations with lower mean reciprocal \ce{Al-N} distances are generally lower in energy. The energy span is larger and correlation clearer for the AAB orientation (\cref{pe_vs_dist} left) than for the AAA.  The results suggest that the ability of cationic quaternary N centers to form close contacts with Al-substituted T-sites is a leading, although not sole, contributor to the potential energy differences, and that those close contacts are more common in the AAA than the AAB orientation of TMAda$^+$. \textcolor{red}{Such observation is also complementary to explain the behavior of the left two panels in Figure 4: in the lower energy region, since AAB can provide wider range of Al-O distances, the AAB energy profile grows slower than that of AAA; while in the higher energy region, the AAB orientation can form longer Al-O distance, which makes AAB orientation have higher energies.}




\subsection{Al Ordering Analysis}
We next explored the relationship between energy and Al configuration in the AAA \ce{TMAda+} subset. As shown in the top left panel in \cref{pe_aaa}, configurations span relative energies between 0 and \SI{40}{kJ/mol_{d6r}}, with a large density of configurations in the intermediate energy regime and sharper variations at the two extremes. By construction, the points include symmetry redundant configurations; for example, by inspection, the six lowest energy configurations are symmetry-equivalent realizations of the same structure. The energy spread of less than \SI{2}{kJ/mol_{d6r}} is within the threshold of energy fluctuation within the block-average sampling method. 
\begin{figure}[tbh]
\centering
\includegraphics[scale=0.5]{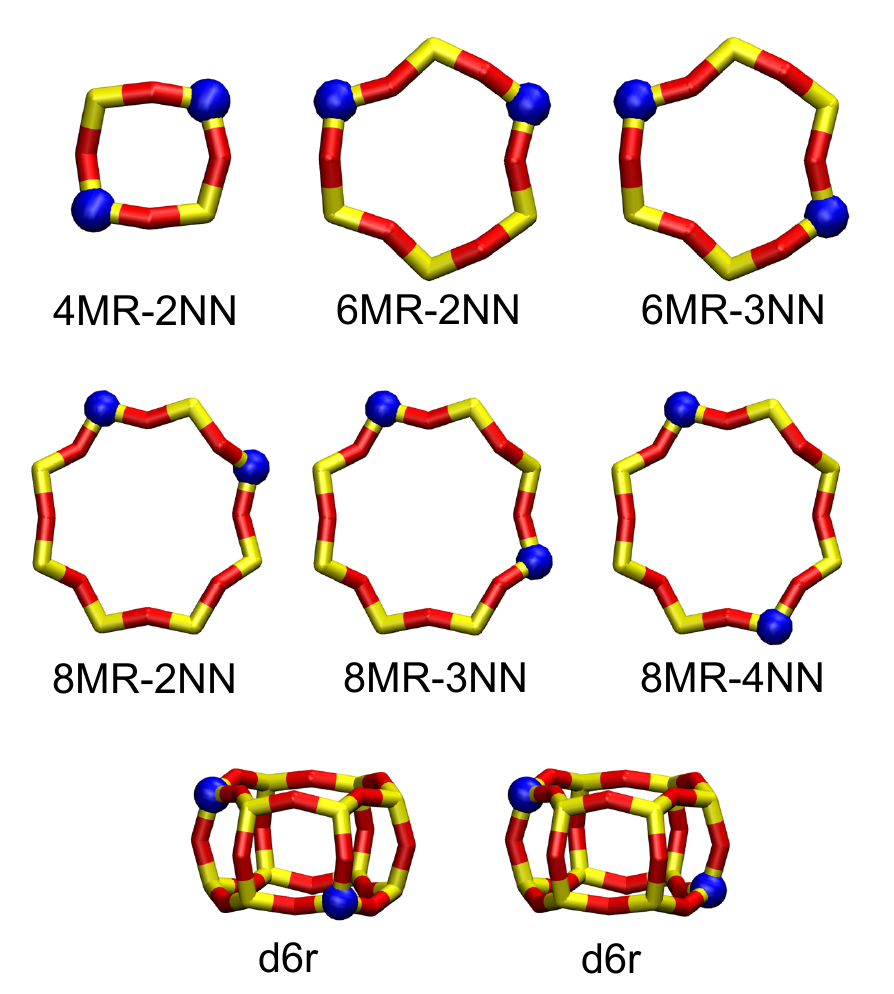}
\caption{\ce{Al-Al} pair proximity features in the CHA zeolite framework. All other pairs are categorized as ``isolated'' pairs.}
\label{ring_type}
\end{figure}

\begin{figure}[tbh]
	\centering
    \includegraphics[scale=0.58]{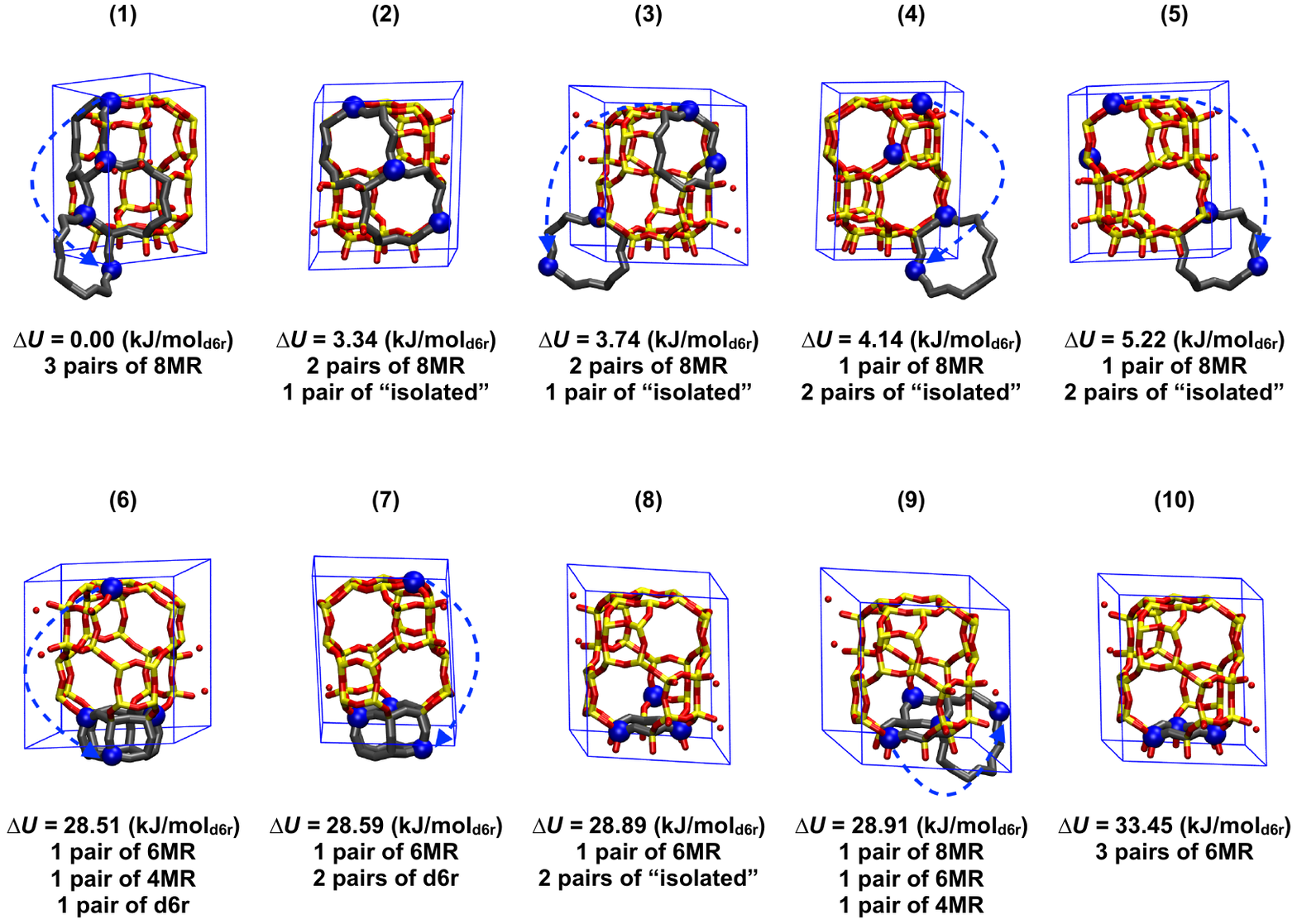}
    \caption{Lowest-energy (1-5) and highest-energy (6-10) AAA CMD supercells and corresponding Al pair features. Features within the supercell are highlighted in gray and features across cell boundary indicated with blue dashed arrows. Color: blue, Al; yellow, Si; red, O.}
	\label{fw_aaa}
\end{figure}

To fingerprint each configuration, we identified the three shortest \ce{Al-Al} contacts and classified each pair either as one of the features shown in \Cref{ring_type} or as an ``isolated'' pair. \textcolor{red}{We want to note here that our definition of ``isolated'' pairs includes all the Al-to-Al structures which do not fall into the category of ``8MR'', ``6MR'', ``4MR'' or ``d6r'' as we listed in \Cref{ring_type}, and such definition has nothing related to the ability to balance divalent cations.} The right panel of \cref{pe_aaa} reports histograms of these pair types vs energy in \SI{2.0}{kJ/mol_{d6r}} wide bins, with histogram points plotted at the low-energy side of the bin. The areas beneath each histogram reflect the relative probabilities of each pair type using the 36 T-site configuration construction algorithm and the assumed 11/1 Si/Al ratio. Isolated pairs are statistically most common, followed by 8MR. Generally, in the low energy region, configurations are rich in 8MR and isolated pairs and poor in 4MR and 6MR pairs. In contrast, high energy configurations contain a mix of features, with 6MR slightly more prominent at the highest energies. \Cref{fw_aaa} shows five non-degenerate representative low- and five non-degenerate representative high-energy Al configurations from the AAA set. All five low energy Al configurations have Al pairs on an 8MR, while the five high energy Al configurations have Al pairs on the d6r unit, 6MR or 4MR. This placement on the 8MR appears to maximize the close contacts with the quaternary ammonium center of \ce{TMAda+}. AIMD results agree with CMD predictions of the large energy difference between the low and high energy Al configurations. AIMD and CMD predictions do not agree on the precise identity of the lowest energy configuration. Among all AIMD-computed configurations, AIMD predicts a configuration that contains only isolated Al as the lowest average potential energy (\SI{2.0}{kJ/mol_{d6r}} lower in energy than the lowest energy configuration in structures 1-5 of \cref{fw_aaa}). AIMD and CMD predict the same highest energy configuration (structure 10 of \cref{fw_aaa}).   

To uncover patterns in energy vs features, we separated the configurations into three subsets by fitting the energy vs configuration data to two 3rd order polynomials across the first and second halves of the profile and partitioning at the two inflection points.
This partitioning resulted in 1499, 3098, and 311 configurations in the low, medium, and high energy bins.  We then Boltzmann-weighted the configurations within each bin, arbitrarily choosing \SI{433}{K} because it is representative of typical zeolite synthesis conditions. $RT$ at \SI{433}{K} is \SI{3.6}{kJ/mol}, so that averaging captures significant fractions of each bin.
The configurational integral of the system with the AAA TMAda$^+$ orientation is
\begin{equation} \label{partition}
    Z^\text{AAA} = \sum_{i} e^{-\Delta U_i/kT }
\end{equation}
where $i$ denotes an Al configuration, $\Delta U_i$ is defined by \cref{delta_e}, $k$ is the Boltzmann constant, $T$ is the averaging temperature, and only AAA orientations are considered in the summation.  The probability of an Al configuration $i$ with an AAA TMAda$^+$ orientation is
\begin{equation}
\begin{aligned}
P_i & = \frac{e^{-\Delta U_i/kT}}{Z^\text{AAA}} 
\end{aligned}
\end{equation}
Finally, the probability $\Pi_j$ of a particular Al pair feature $j$ is
\begin{equation} \label{sum_probability}
\begin{aligned}
\Pi_j &= \sum_{i} \frac{n_{j,i}}{3} P_i
\end{aligned}
\end{equation}
where the $n_{j,i}$ stands for the number of Al pair types $j$ in Al configuration $i$. The factor of 3 accounts for the fact that there are three Al pairs in each configuration.

\begin{figure}[tbh]
	\centering
    \includegraphics[scale=0.5]{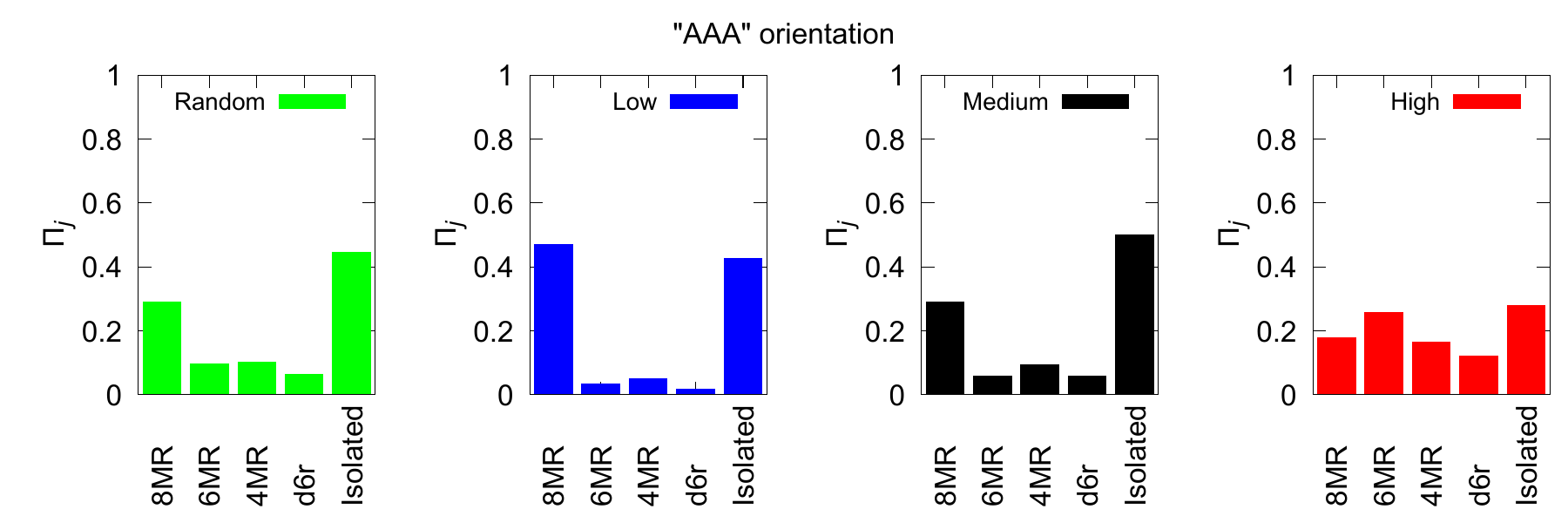}
	\caption{Probability distributions ($\Pi_j$) of the AAA TMAda$^+$ orientations. Left panel (green) represents the random Al distribution with L\"{o}wenstein's rule. ``Low" (blue), ``middle" (black), and ``high" (red) distributions from Boltzmann weightings over configurations subpartitioned by relative energy.}
	\label{probability_aaa}
\end{figure}

\Cref{probability_aaa} compares the probabilities of Al pair types within each energy bin with that expected from a random distribution of Al subject to L\"{o}wenstein's rule, corresponding to the integral of the histograms in \cref{pe_aaa}. Within this random distribution the most probable Al pair types are 8MR and ``isolated'', with a relatively small population of 6MR. The low-energy bin is similarly dominated by 8MR and ``isolated'' features; further, the 6MR and 4MR Al pair features occur at a very low probability. Those 6MR pair features, in contrast, are common in the high energy bin. Results are consistent with the AAA orientation of TMAda$^+$ biasing against Al close contacts.

The bottom left panel in \cref{pe_aaa} reports the mean potential energies of the AAB configurations, plotted using the same reference energy as that for the AAA configurations. Lowest energy configurations are of similar energy, but AAB energies span \SI{60}{kJ/mol_{d6r}}, exceeding the AAA span by more than \SI{20}{kJ/mol_{d6r}}. The bottom right panel in \cref{pe_aaa} shows the corresponding \ce{Al-Al} feature histogram. 6-MR, 4-MR, and d6r Al pairs are represented more prominently at low energy in the AAB orientation than in AAA. 

\begin{figure}[H]
	\centering
    \includegraphics[scale=0.58]{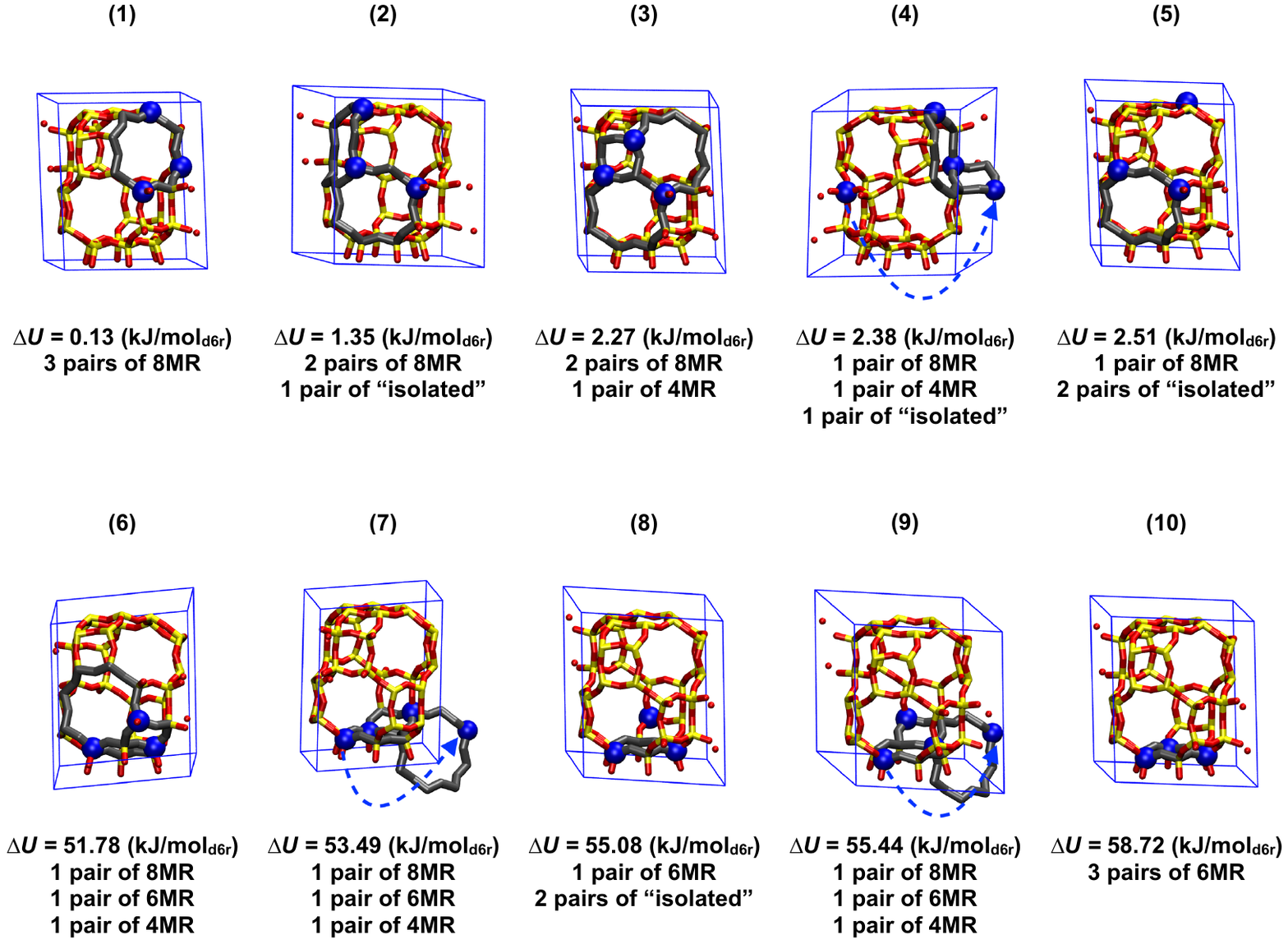}
    \caption{Lowest-energy (1-5) and highest-energy (6-10) AAB CMD supercells and corresponding Al pair features. Features within the supercell are highlighted in gray and features across cell boundary indicated with blue dashed arrows. Color: blue, Al; yellow, Si; red, O.}
	\label{fw_aab}
\end{figure}

\begin{figure}[H]
	\centering
    \includegraphics[scale=0.5]{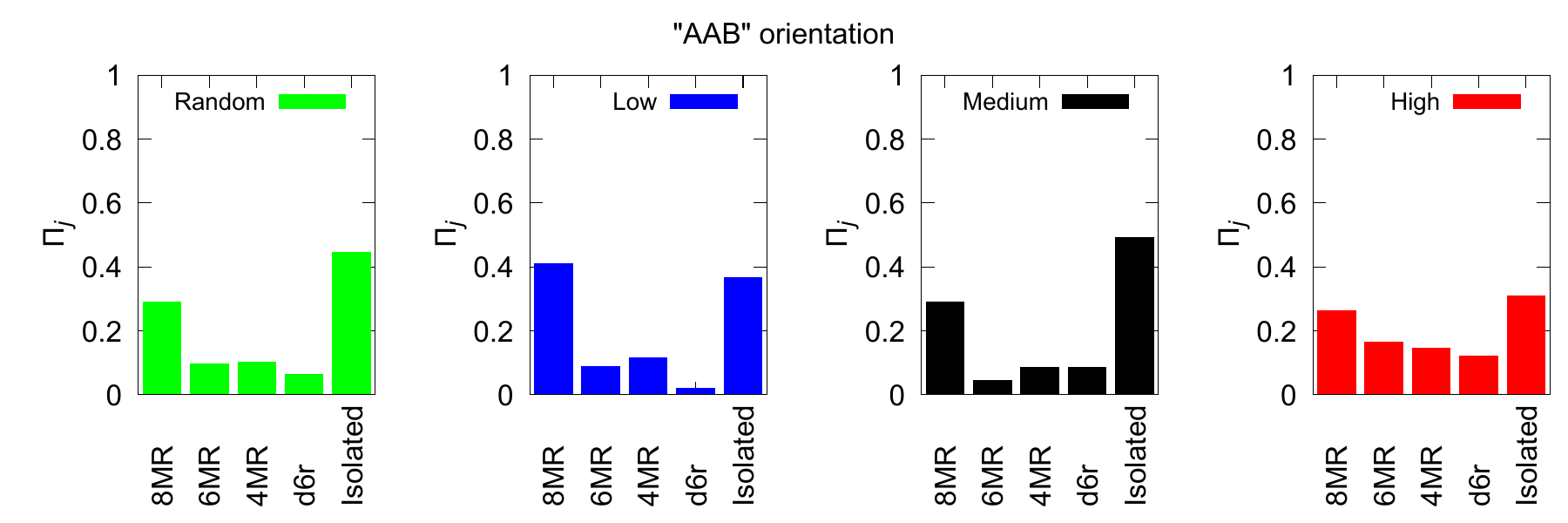}
	\caption{Probability distributions ($\Pi_j$) of the AAB TMAda$^+$ orientations. Left panel (green) represents the random Al distribution with L\"{o}wenstein's rule. ``Low" (blue), ``middle" (black), and ``high" (red) distributions from Boltzmann weightings over configurations subpartitioned by relative energy.}
	\label{probability_aab}
\end{figure}

\Cref{fw_aab} shows snapshots of six lowest and six highest energy AAB configurations, while \cref{probability_aab} shows the probabilities of Al pair types for the AAB orientation, calculated using the same strategy as described above for the AAA orientation (1629, 2920 and 359 Al configurations for low energy, medium energy and high energy region, respectively). Similar to the AAA orientation, all five low energy snapshots have 8-MR Al pair types, and most of these also have isolated Al pairs. But in the AAB orientation, two of the low energy configurations also contain 4-MR pairs. The five high energy configurations all contain 6-MR Al pair types, as observed in the AAA orientation. As with the AAA orientation, AIMD and CMD predictions for AAB agree in terms of gross energy differences but differ in terms of the lowest energy structures. Among all AIMD-computed configurations, AIMD predicts a configuration that contains only isolated Al to have the lowest average potential energy (\SI{2.0}{kJ/mol_{d6r}} lower in energy than the lowest energy configuration in structures 1-5 of \cref{fw_aab}). AIMD and CMD predict the same highest energy configuration (structure 10 of \cref{fw_aab}).

\Cref{probability_aab} shows the Al pair type probabilities computed at \SI{433}{K} for the AAB orientation.
Overall, the probability distribution is similar to the AAA orientation. Frameworks are predicted to be enriched in 8-MR pairs relative to a random distribution and to have a large fraction of isolated pairs. The probability of 6MR and 4MR pairs, however, is comparable to the random distribution and considerably greater than the probability in the AAA orientation. Al pair distributions are thus sensitive to \ce{TMAda+} relative ordering, suggesting a strategy for controlling that distribution. CHA zeolites crystallized with TMAda$^+$ as the sole SDA, with no additional \ce{Na+}, are observed to be poor in 6MR Al pairs \cite{di2016controlling,kester2021effects}, consistent with a large manifold of low energy AAA configurations and the lowest-energy AAB configurations.  

\begin{figure}[H]
\centering
\includegraphics[scale=0.4]{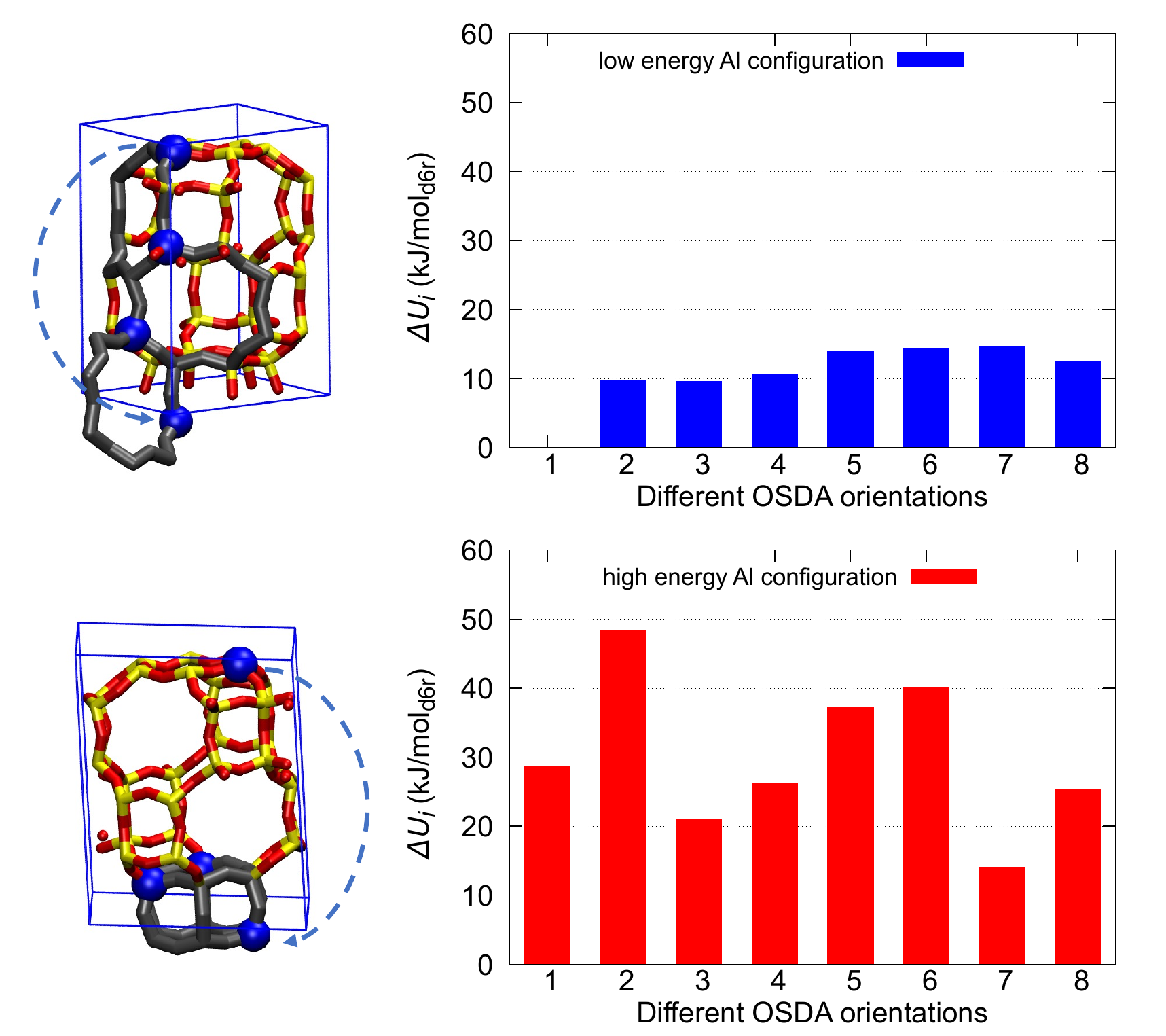}
\caption{Mean potential energies of an 8-MR-pair-only configuration (top) and a configuration containing 6-MR and d6r pairs (bottom), each in the field of all possible TMAda$^+$ orientations.  Al T-sites highlighted in blue and arrows indicate periodic images.}
\label{flip_osda}
\end{figure}

To further explore the sensitivity of potential energy to TMAda$^+$ orientation, we selected the lowest-energy Al configurations from the AAA set, which contains only 8MR pairs, flipped the orientations of each TMAda$^+$ to create eight OSDA orientations, and computed averaged potential energies. Results are shown in \Cref{flip_osda} top. Flipping OSDAs within the low energy configuration results in a number of degenerate structures due to system symmetry. The energy cost to flip TMAda$^+$ is modest (\SI{10}{kJ/mo_{d6r}}) and essentially constant---this 8MR-only structure, which avoids 6MR pairs, is relatively robust to OSDA orientation. We applied the same strategy to an Al configuration that contains a 6MR and d6r pairs. As shown in the bottom of \Cref{flip_osda}, the energy of this configuration is highly sensitive to TMAda$^+$ orientation. While its lowest energy realization is competitive in energy with the 8MR-only structure, most orientations lead to much higher energies. These results suggest that some features, such as the 8-MR pair, may be preferentially biased for because they are agnostic to local OSDA orientation, while others, such as the 6-MR pair, are biased against because they are more sensitive to local OSDA orientation. 

\begin{figure}[H]
\centering
\includegraphics[scale=0.6]{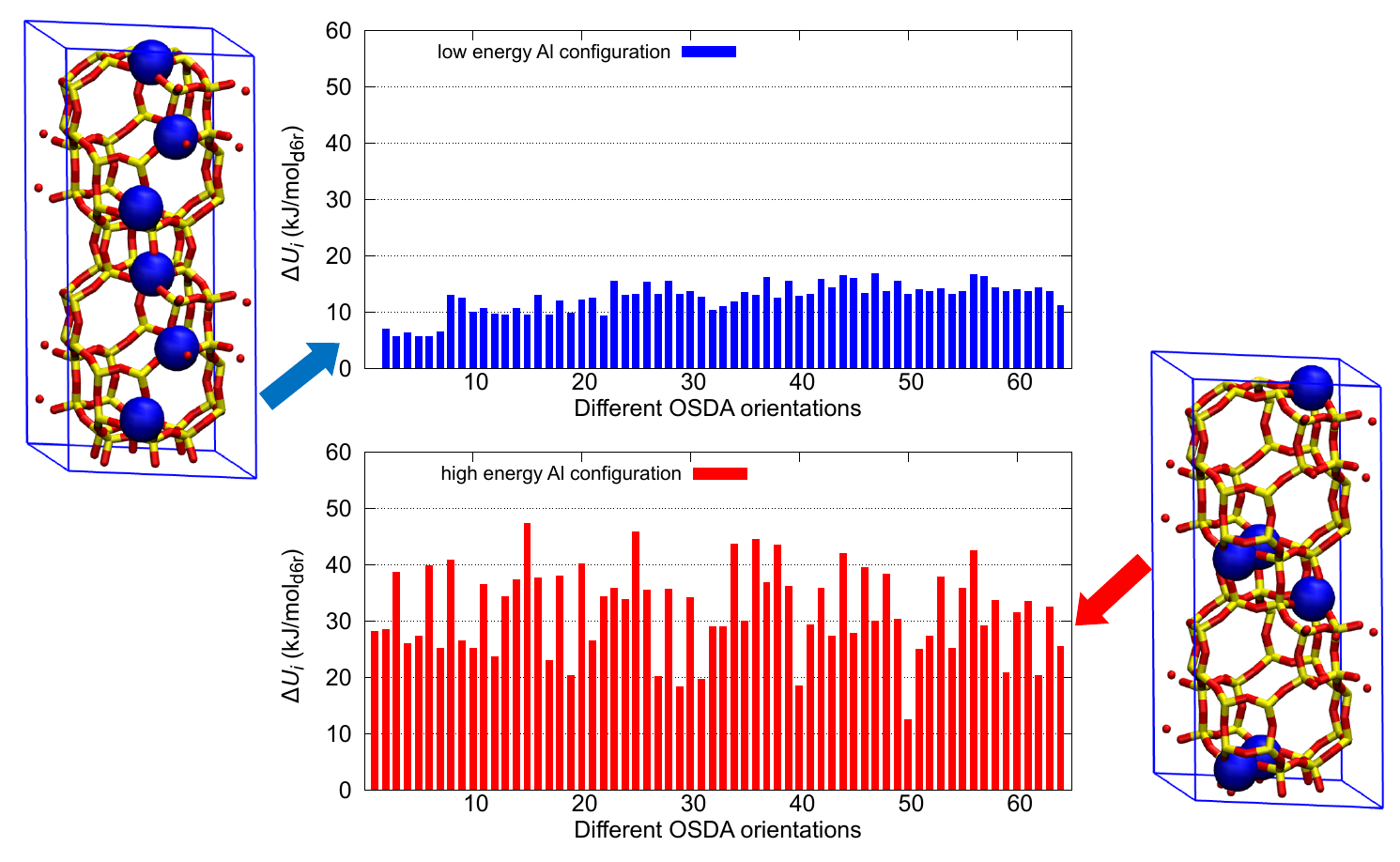}
\caption{Mean potential energies of a 72-T-site 8-MR-pair-only configuration (top) and a 72-T-site configuration containing 6-MR and d6r pairs (bottom), each in the field of all possible TMAda$^+$ orientations. Same OSDA orientation with different Al configurations are stacked. Al T-sites highlighted in blue.}
\label{pot_72t}
\end{figure}

We doubled the supercells shown in \Cref{flip_osda} along the $c$ direction and repeated the TMAda$^+$ flipping procedure, creating $64$ OSDA orientational combinations per supercell. Supercells and mean potential energies are shown in \cref{pot_72t}. Results mirror those of \Cref{flip_osda}: the 8-MR-only Al configuration is minimized in energy when all TMAda$^+$ are aligned in the same orientation (``AAAAAA'') but energy costs to ``flip'' OSDA are relatively small and constant. In contrast, the 6-MR Al configuration is higher in energy across all configurations save number 50; further, energies are much more sensitive to \ce{TMAda+} orientations. 

\section{Conclusions}
While the ability to guide zeolite structure through OSDA selection, and approaches to simulate this influence, are well established, the relationship between OSDA choice and the distribution of Al on a zeolite framework is less clear. CHA is an ideal template for exploring these effects, as it has only one symmetry-distinct T-site; all T-sites thus see identical void environments within the framework, and non-random Al distributions must reflect either kinetic or thermodynamic factors at play during crystallization. Here we explore the thermodynamics of Al distributions in the field of TMAda$^+$ structure directing agents, using a combination of \textit{ab initio} and classical dynamics models of TMAda$^+$ occluded within the three cages of a CHA unit cell.  System energies are observed to be sensitive both to Al proximity and to the field of TMAda$^+$, and energy variations are consistent with a substantial contribution from OSDA-Al electrostatics, related to the ability of quaternary nitrogen to approach Al T-sites. Configurations the place Al pairs in 8MRs maximize favorable electrostatic contacts with \ce{TMAda+} and are low in energy. Further, the energies of those configurations are less sensitive to \ce{TMAda+} orientation than are Al pairs in smaller rings. This robustness to OSDA order (or disorder) may thus be a relevant factor determining Al distributions.  
Al configurations that place two Al within the same 6MR are high in energy, consistent with experimental observations that these features are rare on CHA zeolite prepared with TMAda$^+$ as the sole structure-directing agent \cite{di2016controlling,di2017}. This correspondence suggests that lattice energies are, at least in this system, a useful predictor of Al siting preferences, as has been observed in similar simulations exploring the influence of \ce{Na+} and \ce{TMAda+} co-occlusion on Al siting \cite{di2020cooperative}. 

The results highlight the potential to apply similar strategies to other OSDAs and frameworks. The CHA-TMAda$^+$ system is simplified by the fact that \ce{TMAda+} can adopt only one of two primary orientations within the cha cage, and models here are limited to Al as the sole charge-carrying site on the framework. Further extensions will benefit from improvements in forcefield parameterization, in configurational sampling, and in model generalizations to framework compositions away from 1:1 OSDA to Al. 

\begin{acknowledgement}
The authors gratefully acknowledge financial support from the National Science Foundation CBET-DMREF program under award number 1922154. The computing resources for this work were provided by the Notre Dame Center for Research Computing.
\end{acknowledgement}

\begin{suppinfo}
A PDF file is provided that contains:
\begin{itemize}
  \item CMD, AIMD energies and uncertainties plotted in Figure \ref{AIMD_CMD}.
  \item Analysis of data scattering in Figure \ref{AIMD_CMD}.
  \item Analysis of the correlations of the Al-Al distance and N-N distance with the relative potential energy ($\Delta U_i$).
\end{itemize}
 
A ZIP file (link: https://doi.org/10.5281/zenodo.6422186) is provided that contains:
\begin{itemize}
  \item VASP input files for geometry optimization, charge generation, and AIMD simulations.
  \item LAMMPS input file for CMD simulations.
  \item CONTCARs of DFT optimized structures used for charge analysis and corresponding XYZ files that contain raw partial charges.
  \item POSCARs of 72 structures that are used for the analysis of CMD and AIMD correlations.
  \item TXT files containing raw data that is used to plot figures.
  \item \textcolor{red}{XYZ files of 4908 ``AAA'' structures simulated by CMD.}
  \item \textcolor{red}{XYZ files of 4908 ``AAB'' structures simulated by CMD.}
\end{itemize}

\end{suppinfo}

\bibliography{references}

\end{document}